\documentclass{elsart3p}

\usepackage{graphicx}
\usepackage{graphics}
\usepackage{amssymb,amsmath}

\begin{document}

\begin{frontmatter}



\title{Thermally induced 0-$\pi$ phase transition in Josephson junctions through a ferromagnetic oxide film}


\author[address1,address2]{S. Kawabata},
\author[address3]{Y. Asano}, 
\author[address4]{A. A.~Golubov},
\author[address5]{Y. Tanaka},
\author[address6]{S. Kashiwaya}

\address[address1]{Nanosystem Research Institute (NRI), National Institute of Advanced Industrial Science and Technology (AIST), Tsukuba, Ibaraki 305-8568, Japan}

\address[address2]{CREST, Japan Science and Technology Corporation (JST), Kawaguchi, Saitama 332-0012, Japan}

\address[address3]{Department of Applied Physics, Hokkaido University, Sapporo, 060-8628, Japan}

\address[address4]{Faculty of Science and Technology, University of Twente, P.O. Box 217, 7500 AE Enschede, The Netherlands}

\address[address5]{Department of Applied Physics, Nagoya University, Nagoya, 464-8603, Japan}

\address[address6]{Nanoelectronics Research Institute (NeRI), AIST, Tsukuba, Ibaraki 305-8568, Japan}



\begin{abstract}

We investigate the Josephson transport through a  ferromagnetic oxide film, e.g., La$_2$BaCuO$_5$, theoretically.
Using the recursive Green's function technique, we found the formation of a $\pi$-junction in such systems.
 Moreover the 0-$\pi$ phase transition is induced by increasing the temperature. 
Such ferromagnetic-oxide based Josephson junctions may become an element in the architecture of future quantum
computers.
\end{abstract}

\end{frontmatter}

\newpage
\section{Introduction}

It is  well established that a superconducting phase difference of $\pi$ can be arranged between two $s$-wave superconductors (Ss) when separating them by a suitably chosen ferromagnetic metals (FMs)~\cite{rf:Buzdin,rf:Golubov}. 
Transitions between the $\pi$-state and the 0-state of such S/FM/S Josephson junctions have been revealed in experiments through oscillations of the Josephson critical current $I_c$ with varying thickness of the FM~\cite{rf:Kontos} or with varying temperature~\cite{rf:Ryanzanov}. 
The $\pi$ Josephson junction is currently of considerable interest as an element complementary to the usual Josephson junction in the development of functional nano-structures including superconducting electronics~\cite{rf:Ortlepp}.

Recently, $quiet$ qubits consisting of a S/FM/S $\pi$ junction have been theoretically proposed~\cite{rf:Ioffe,rf:Blatter}.
In the quiet qubits, a quantum two level system is spontaneously generated and therefore it is expected to be robust to the decoherence by the fluctuation of the external magnetic field.
From the viewpoint of the quantum decoherence, however, S/FM/S junctions is inherently identical with S/NM/S junctions, where NM is a nonmagnetic normal-metal.
Therefore a gapless quasiparticle excitation in the FM region is inevitable.
This feature gives a strong Ohmic dissipation~\cite{rf:Zaikin} and the coherence time of S/FM/S quiet qubits is bound to be very short.
Thus the realization of the $\pi$ junction $without$ using such metallic ferromagnets are highly desired for qubit applications~\cite{rf:Kawabata1,rf:Kawabata2,rf:Kawabata3,rf:Kawabata4}.

On the other hand, recently, we have developed a  numerical method to calculate the Josephson current by taking into account the band structure of ferromagnetic materials based on the Recursive Green's function method~\cite{rf:Kawabata5,rf:Kawabata6,rf:Kawabata7}.
By use of this method, we numerically found the formation of the $\pi$-state for the $s$-wave junction through a ferromagnetic oxide (FOs) which is a kind of ferromagnetic insulators.
Moreover we have found that the 0-$\pi$ transition is induced by increasing the thickness of the oxide layer~\cite{rf:Kawabata6,rf:Kawabata7,rf:Kawabata8}. 
Heretofore, however, we have only considered the Josephson transport at the very low temperature regime, i.e., $T \ll T_c$, where $T_c$ is the superconductor transition temperature.
In this paper we will investigate the $finite$-temperature Josephson transport and show that the 0-$\pi$ phase transition can be realized by increasing the $temperature$.

\section{Electronic state of ferromagnetic oxides}
\begin{figure}[b]
\begin{center}
\includegraphics[width=8.0cm]{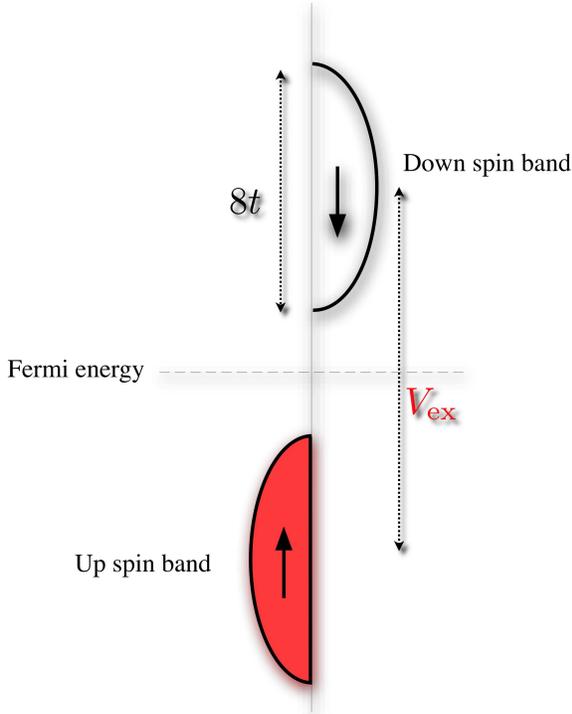}
\end{center}
\caption{(Color online) The density of states for each spin direction for a ferromagnetic oxide, e.g.,  La${}_2$BaCuO${}_5$.}
\label{fig1}
\end{figure}

The typical DOS of FOs for each spin direction is shown schematically in Fig. 1.
One of the representative material of such magnetic materials is half-filled La${}_2$BaCuO${}_5$ (LBCO)~\cite{rf:Mizuno,rf:Masuda,rf:Ku}.
The exchange splitting $V_\mathrm{ex}$ is estimated  to be 0.34 eV by a first-principle band calculation using the spin-polarized local density approximation~\cite{rf:LBCO}.
Since the exchange splitting $V_\mathrm{ex}$ is large and the bands are originally half-filled, the system becomes  the ferromagnetic insulator. 
In the next section, we calculate the Josephson current through such oxide films numerically.

\section{Numerical calculation of Josephson current}

In this section, we outline a numerical calculation method for the Josephson current of S/FO/S junctions based on the recursive Green's function technique~\cite{rf:Asano}.
Let us consider a two-dimensional square tight-binding lattice for the S/FO/S junction.
The vector $\boldsymbol{r}=j{\boldsymbol{x}}
+m{\boldsymbol{y}}$ points to a lattice site, where ${\boldsymbol{x}}$ and ${\boldsymbol{y}}$ are unit vectors in the $x$ and $y$ directions,
respectively.
In the $y$ direction, we apply the periodic boundary condition for the number of lattice sites being $W$.

Electronic states in a superconductor are described by the
mean-field Hamiltonian
 \begin{eqnarray}
H_{\mathrm{BCS}}&=& \frac{1}{2}\sum_{\boldsymbol{r},\boldsymbol{r}^{\prime }  \in \text{S}    }%
\left( \tilde{c}_{\boldsymbol{r}}^{\dagger }\;\hat{h}_{\boldsymbol{r},%
\boldsymbol{r}^{\prime }}\;\tilde{c}_{\boldsymbol{r}^{\prime }}^{{}}-%
\overline{\tilde{c}_{\boldsymbol{r}}}\;\hat{h}_{\boldsymbol{r},\boldsymbol{r}%
^{\prime }}^{\ast }\;\overline{\tilde{c}_{\boldsymbol{r}^{\prime }}^{\dagger
}}\;\right)
\nonumber\\
 &+&
 \frac{1}{2}\sum_{\boldsymbol{r}\in \text{S}}\left( \tilde{c}_{%
\boldsymbol{r}}^{\dagger }\;\hat{\Delta}\;\overline{\tilde{c}_{\boldsymbol{r}%
}^{\dagger }}-\overline{\tilde{c}_{\boldsymbol{r}}}\;\hat{\Delta}^{\ast }\;%
\tilde{c}_{\boldsymbol{r}}\right) 
.
\end{eqnarray}
In this equation,
 \begin{eqnarray}
\hat{h}_{\boldsymbol{r},\boldsymbol{r}^{\prime }}&=& \left\{ -t_s \delta _{|%
\boldsymbol{r}-\boldsymbol{r}^{\prime }|,1}+(-\mu_s
+4t_s)\delta _{\boldsymbol{r},\boldsymbol{r}^{\prime }}\right\} \hat{\sigma}_{0}
,
\end{eqnarray}
with 
$\overline{\tilde{c}}_{\boldsymbol{r}}=\left( c_{\boldsymbol{r}%
,\uparrow },c_{\boldsymbol{r},\downarrow }\right) $,
 where
  $
  c_{\boldsymbol{r} ,\sigma }^{\dagger }$ ($c_{\boldsymbol{r},\sigma }^{{}}
$)
 is the creation
(annihilation) operator of an electron at $\boldsymbol{r}$ with spin $\sigma
=$ ( $\uparrow $ or $\downarrow $ ), $\overline{\tilde{c}}$ means the
transpose of $\tilde{c}$,  and $\hat{\sigma}_{0}$ is $2\times 2$ unit matrix. 
The chemical potential  $\mu_s$ is set to be $2 t_s$ for superconductors.
In superconductors, the hopping integral $t_s$ is considered among nearest neighbor sites and we choose 
 \begin{eqnarray}
\hat{\Delta}=i\Delta \hat{\sigma}_{y}
,
\end{eqnarray}
 where $\Delta $ is the amplitude 
of the pair potential in the $s$-wave symmetry channel, and $\hat{\sigma}_{y}$ is a Pauli matrix.

The Hamiltonian of the ferromagnetic oxide film is given by a single-band tight-binding model as 
 \begin{eqnarray}
H_\mathrm{FO} &=& -t \sum_{\boldsymbol{r},\boldsymbol{r}^{\prime },\sigma} 
c_{\boldsymbol{r},\sigma}^\dagger 
c_{\boldsymbol{r}',\sigma}
-\sum_{\boldsymbol{r}} ( 4 t -\mu)  
c_{\boldsymbol{r},\uparrow}^\dagger 
c_{\boldsymbol{r},\uparrow}
\nonumber\\
&+&
 \sum_{\boldsymbol{r}} 
( 4 t -\mu + V_\mathrm{ex}) 
 c_{\boldsymbol{r},\downarrow}^\dagger 
 c_{\boldsymbol{r},\downarrow}
,
  \end{eqnarray}
where $V_\mathrm{ex}$ is the exchange splitting between the up and down spin band.
When $V_\mathrm{ex} >  8 t$, this Hamiltonian describes the ferromagnetic oxide as shown in Fig. 1.
The chemical potential $\mu$ is set to be $V_\mathrm{ex}/2  + 4t$. 
A superconductor and a ferromagnetic oxide are connected by 
\begin{eqnarray}
H_{c1} = &-& t_s \sum_{m,\sigma} \left( c_{0,m,\sigma}^\dagger c_{1,m,\sigma} + c_{L_F,m,\sigma}^\dagger c_{L_F+1,m,\sigma} \right.
\nonumber\\
&+& \left. h. c. \right)
,
\end{eqnarray}
where $L_F$ is the thickness of the FO layer.

The Hamiltonian is diagonalized by the Bogoliubov transformation and the
Bogoliubov-de Gennes equation is numerically solved by the recursive
Green function method\cite{rf:Asano}.
We calculate the Matsubara
Green function in a ferromagnetic oxide,
\begin{equation}
\check{G}_{\omega _{n}}(\boldsymbol{r},\boldsymbol{r}^{\prime })=\left(
\begin{array}{cc}
\hat{g}_{\omega _{n}}(\boldsymbol{r},\boldsymbol{r}^{\prime }) & \hat{f}%
_{\omega _{n}}(\boldsymbol{r},\boldsymbol{r}^{\prime }) \\
-\hat{f}_{\omega _{n}}^{\ast }(\boldsymbol{r},\boldsymbol{r}^{\prime }) & -%
\hat{g}_{\omega _{n}}^{\ast }(\boldsymbol{r},\boldsymbol{r}^{\prime })%
\end{array}
\right) , \label{deff}
\end{equation}
where $\omega _{n}=(2n+1)\pi T$ is the Matsubara frequency, $n$ is an
integer number, and $T$ is a temperature. 
The finite temperature Josephson current is given by
\begin{equation}
I_J (\phi)=-ietT\sum_{\omega _{n}}\sum_{m=1}^{W}\mathrm{Tr}\left[ \check{G}_{\omega
_{n}}(\boldsymbol{r}^{\prime },\boldsymbol{r})-\check{G}_{\omega _{n}}(%
\boldsymbol{r},\boldsymbol{r}^{\prime })\right]
,
\end{equation}
with $\boldsymbol{r}^{\prime }=\boldsymbol{r}+\boldsymbol{x}$. 
Note that the Green function in Eq.~(\ref{deff}) 
is a $4\times 4$ matrix representing spin and Nambu spaces. 
Throughout this paper we fix the following
parameters: $W=5$, and $\Delta _{0}=0.01t$, and assume $t=t_s$ for simplicity.

\section{Result}
\begin{figure}[t]
\begin{center}
\includegraphics[width=7.5cm]{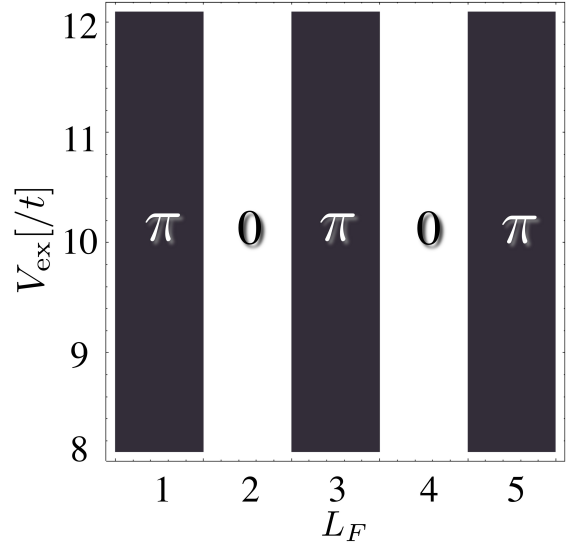}
\end{center}
\caption{The phase diagram depending on the exchange splitting  $V_\mathrm{ex}$ and the thickness of the ferromagnetic oxide (FO) layer $L_F$ for the  S/FO/S Josephson junction at $T=0.01 T_c$.
The black and white regime correspond to the $\pi$ and 0 junction, respectively. }
\label{fig2}
\end{figure}

The 0-$\pi$ phase diagram for $T=0.01T_c$ depending on the exchange splitting $V_\mathrm{ex}$  and the thickness of FO layer $L_F$ is shown in Fig.~\ref{fig2}.
The black (white) regime corresponds to the $\pi$- (0-)junction, i.e., $I_J=  -(+)I_C \sin \phi$.
We find that the atomic-scale 0-$\pi$ phase transition is induced by increasing the thickness of FO layer~\cite{rf:Kawabata6,rf:Kawabata7}.

Next we consider the finite-temperature Josephson transport.
 The temperature dependence of $I_c$ for the odd  $L_F$ junction is plotted in Fig. 3. 
 In the case of the tunneling limit ($V_\mathrm{ex} \gg  t$), $| I_c |$ obeys the well-known Ambegaokar-Baratoff formula. 
 By decreasing the exchange splitting $V_\mathrm{ex}$,  we found the appearance of an anomalous thermally-induced 0-$\pi$
transition for the S/FO/S  junction.
More detailed discussion for above peculiar results will be given elsewhere.

\begin{figure}[t]
\begin{center}
\includegraphics[width=8.0cm]{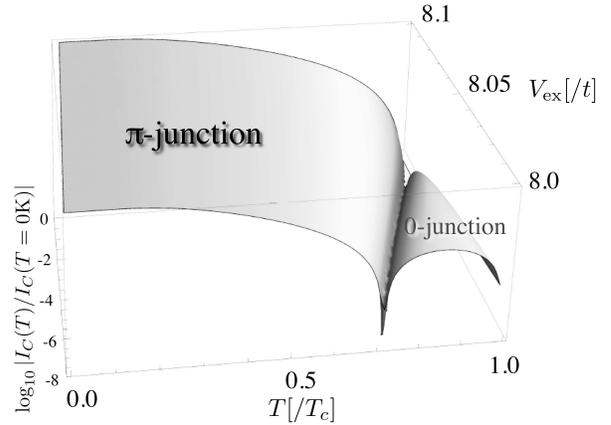}
\end{center}
\caption{
The Josephson critical  current $I_C$ as a function of the temperature $T$ and the  exchange splitting  $V_\mathrm{ex}$ for  the  S/FO/S Josephson junction with $L_F$=3.
}
\label{fig3}
\end{figure}
\
\section{Summary}
To summarize, we have studied the finite-temperature Josephson transport in S/ferromagnetic-oxide/S junction by use of the recursive Green's function method.
We found that the 0-$\pi$ transition can be induced by increasing the temperature.
Therefore we can experimentally confirm the $\pi$ junction behavior by  measuring the temperature dependence of the Josephson critical current. 
Such FO junctions may become an element in the architecture of  $\pi$-junction based qubits~\cite{rf:Kawabata1,rf:Kawabata2,rf:Kawabata3,rf:Kawabata4}.

\section*{Acknowledgements}

This work was  supported by CREST-JST, and a Grant-in-Aid for Scientific Research from the Ministry of Education, Science, Sports and Culture of Japan (Grant No. 22710096).

\end{document}